\documentclass[aps,prd,twocolumn]{revtex4-2}

\usepackage{amssymb}
\usepackage{amsmath}

\usepackage{graphics}
\usepackage[export]{adjustbox}

\usepackage{xparse}

\usepackage{hyperref}
\usepackage{hypernat}

\usepackage{layouts}

\usepackage[caption=false,labelfont=normalsize,labelformat=empty, textfont=normalsize,justification=centering]{subfig}

\begin{document}

\title{Comment on ``Dirac leptogenesis via scatterings''}

\author{Tom\'a\v s Bla\v zek}
\email{tomas.blazek@fmph.uniba.sk}
\author{Peter Mat\'{a}k}
\email{peter.matak@fmph.uniba.sk}
\affiliation{Department of Theoretical Physics, Comenius University,\\ 
Mlynsk\'a dolina, 84248 Bratislava, Slovak Republic}

\date{\today}

\begin{abstract}
In their recent letter, Heeck {\it et al.} [\href{https://link.aps.org/doi/10.1103/PhysRevD.108.L031703}{Phys. Rev. D {\bf 108}, L031703}]\nocite{Heeck:2023rrz} introduced a simple model of leptogenesis, in which the asymmetry originates from the scattering processes only. Here we argue that their calculation violates some of the basic principles of perturbative quantum field theory. 
\end{abstract}
\maketitle

The $CPT$ and unitarity constraints \cite{Kolb:1979qa, Dolgov:1979mz, Baldes:2014gca} require the $CP$ asymmetries with a fixed initial state to cancel when summing over all possible final states. For generic $i\rightarrow f$ reactions, the asymmetries of squared amplitudes then obey the relation
\begin{align}\label{eq1}
\sum_f\Delta\vert M_{fi}\vert^2=0.
\end{align}
Let us consider the $e_L \nu_L$ initial state within the model defined in Eq. (1) of Ref. \cite{Heeck:2023rrz}. Unless producing the scalar mediators on-shell, at the given perturbative order, the only available final states contain $e_L \nu_L$ or $e_R\nu_R$ particle pairs. Due to the $CPT$ invariance, the flavour-blind asymmetry of the $e_L\nu_L\rightarrow e_L\nu_L$ process must vanish. Therefore, in consequence of Eq. \eqref{eq1}, and contrary to the result in Eq. (14) in Ref. \cite{Heeck:2023rrz}, the asymmetry of the $e_L\nu_L\rightarrow e_R\nu_R$ reaction must vanish as well. 

To make the statement more explicit, let us look at the tree and $1$-loop $s$-channel diagrams
\begin{align}\label{eq2}
\includegraphics[scale=1,valign=c]{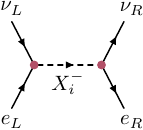}\hskip2mm, \hskip2mm\includegraphics[scale=1,valign=c]{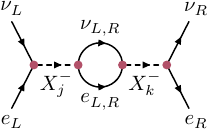}\hskip1mm, 
\end{align}
where we only renamed $X$-field indices compared to the diagrams in the upper part of Fig. 1 in Ref. \cite{Heeck:2023rrz}. The contribution of Eq. \eqref{eq2} to the asymmetry of the process is proportional to the imaginary part of  $\mathcal{F}_{ji}\mathcal{G}_{ik}(\mathcal{F}_{kj}+\mathcal{G}_{kj})$, using the hermitian matrices defined below Eq. (13) of Ref. \cite{Heeck:2023rrz}. Putting $i=2$ and $j=k=1$, the first term in Eq. (14) of Ref. \cite{Heeck:2023rrz} is reproduced. In Ref. \cite{Heeck:2023rrz}, the two other possibilities, $j=2$, $k=i=1$, and $k=2$, $i=j=1$, seem to be missed. By including them properly, we obtain
\begin{align}\label{eq3}
\mathcal{F}_{11} \big(\mathcal{F}_{12}\mathcal{G}_{21}+\mathcal{G}_{12}\mathcal{F}_{21}+ \mathcal{G}_{12}\mathcal{G}_{21}\big)+\big(\mathcal{F}\leftrightarrow \mathcal{G}\big).
\end{align}
This combination of couplings is always real, and no asymmetry in the $e_L \nu_L \rightarrow e_R\nu_R$ cross section can be produced.

Furthermore, apart from irreducible complex phases in couplings, nonzero asymmetry requires imaginary kinematics from loops. That can be represented by Cutkosky \cite{Cutkosky:1960sp} or holomorphic \cite{Coster:1970jy, Bourjaily:2020wvq, Blazek:2021olf} cuts in the respective diagrams. For the $t$-channel loop in the lower part in Fig. 1 in Ref. \cite{Heeck:2023rrz}, no kinematically-allowed cuts are possible. Consequently, at the perturbative order of interest, no asymmetry can occur for the $e_R\Bar{e}_L\rightarrow \Bar{\nu}_R\nu_L$ cross section.

The authors of Ref. \cite{Heeck:2023rrz} do not present a detailed derivation of the results in their Eqs. (14) and (16). However, based on the above reasoning, it appears their treatment is inconsistent with unitarity and $CPT$ invariance. When evaluated correctly, both $\epsilon_s$ and $\epsilon_t$ parameters vanish, and the model in its minimal version cannot explain the matter asymmetry of the universe.

Finally, we also comment on the symmetric part of the cross sections in Eqs. (15) and (17) of Ref. \cite{Heeck:2023rrz}. The presented expressions seem to contain squares of the tree-level diagrams mediated by $X_1$ and analogous contributions of $X_2$. However, the terms in which the $X_1$ and $X_2$ diagrams interfere seem to be missing, rendering the result incomplete.

This work was supported by Slovak Grant Agency VEGA, project No. 1/0719/23.

\bibliographystyle{apsrev4-1.bst}
\bibliography{COMMENT.bib}

\end{document}